\documentclass[traditabstract]{aa}

\usepackage{graphicx}
\usepackage{txfonts}
\usepackage{hyperref}

\usepackage{subcaption}
\authorrunning{I\l{}kiewicz et al.}
\begin{document}

\title{Circumbinary disc interactions and stochastic dust obscuration in the post-asymptotic-giant-branch binary HD 213985}
\titlerunning{Circumbinary disc interactions and stochastic dust obscuration in HD 213985}

\author{Krystian I\l{}kiewicz  \inst{1, }\inst{2}
          \and  L\'ea Planquart\inst{3}
          \and Tomek Kami\'nski \inst{4}
          \and Hans Van Winckel \inst{5}
          }

   \institute{Nicolaus Copernicus Astronomical Center, Polish Academy of Sciences, Bartycka 18, 00-716 Warsaw, Poland\\ \email{ilkiewicz@camk.edu.pl}
     \and
     Astronomical Observatory, University of Warsaw, Al. Ujazdowskie 4, 00-478 Warszawa, Poland
    \and
    Institut d’Astronomie et d’Astrophysique, Universit\'e Libre de Bruxelles, CP 226, Boulevard du Triomphe, 1050 Brussels, Belgium
    \and
    Nicolaus Copernicus Astronomical Center, Polish Academy of Sciences, Rabia{\'n}ska 8, 87-100 Toru{\'n}, Poland
    \and
    Institute of Astronomy, KU Leuven, Celestĳnenlaan 200D, 3001 Leuven, Belgium
             }

   \date{}

  \abstract
  { 
   \object{HD 213985} is an eccentric binary system with a post-AGB primary and a jet-launching secondary star. We confirm that the system photometric variability is likely due to obscuration by the inner edge of a circumbinary disc, similar to RVb-type RV~Tau stars. The system has shown an increase in the orbital variability amplitude in optical photometric bands, along with irregular changes in its shape that often started to appear skewed.  Variability in the Na~D lines suggests that this behaviour may be driven by interactions between the circumbinary disc and outflows through the L2 Lagrange point. Moreover, \object{HD 213985} has exhibited episodes of short-term fluctuations whose appearance is not strictly related to the orbital phase. This variability is consistent with obscuration by transient dust structure leading to weather-like variability patterns.
}


   \maketitle

\section{Introduction}

Post-asymptotic giant branch (post-AGB) objects are low-mass stars in a fast transition phase between the AGB and planetary nebula stages of evolution \citep[e.g.][]{2016A&A...588A..25M}. Post-AGB objects in binaries are of particular interest, as they are important for our understanding of the influence of binarity on stellar evolution \citep{VanWinckel_2016,2018arXiv180900871V}. Binary post-AGB stars often have eccentric orbits and orbital periods on the order of years \citep{2018A&A...620A..85O}. The system configuration typically consists of a post-AGB primary interacting with a jet-launching main sequence companion surrounded by  an accretion disc (hereafter referred to as circumcompanion disc). Moreover, the binaries are surrounded by circumbinary discs ejected by the primary at the end of the AGB stage of evolution \citep[e.g.]{2023A&A...674A.151C}. Roughly 85 post-AGB binaries have been discovered thus far \citep{2022A&A...658A..36K}.

\object{HD 213985} (=HM Aqr) is a binary system with a post-AGB primary \citep{1987A&A...181L...5W,1995Ap&SS.224..357W,2000IAUS..177..285V,2009A&A...498..489J}. The post-AGB star spectroscopic classification is A0III \citep{1988mcts.book.....H}. The infrared spectral energy distribution indicates the presence of a full circumbinary disc, according to the classification of \citet{2022A&A...658A..36K}. The binary period is 259.6$\pm$0.7 days, and the orbit eccentricity is 0.21$\pm$0.05 \citep{2018A&A...620A..85O}. The orbit inclination is $i$=50\fdg1$\pm$0\fdg3 \citep{2022A&A...666A..40B}, and the circumbinary disc is in, or close to, the orbital plane ($i\simeq$60\degr; \citealt{2019A&A...631A.108K}). The accretor in \object{HD 213985} is launching a wide jet, typical for such binaries \citep{2022A&A...666A..40B}. The jet is tilted by 9\fdg9$\pm$0\fdg2 from the binary orbit \citep{2022A&A...666A..40B}.

Since the mid-1980s to early-1990s \object{HD 213985} has shown sinusoidal orbital variability with an amplitude of $\sim$0.1~mag in optical and near-infrared filters \citep{1987A&A...181L...5W,1989MNRAS.241..393W,2007AN....328..848S}. Small changes of the light curve shape between orbital cycles have been noted by \citet{2007MNRAS.375.1338K} in the $V$ band, that the authors have attributed to on-going changes in the circumstellar shell. This is similar to RVb phenomenon in RV~Tau type variables, where the amplitude of pulsating post-AGB star changes due to periodic obscuration by the circumbinary dusty disc \citep{2017A&A...608A..99K}. 

Here we investigate the evolution of variability of \object{HD 213985}. For this purpose, we employed ground-based and space-based photometric observations, as well as high-resolution spectroscopy introduced in Section~\ref{sec:data}. In Section~\ref{sec:results}, we present the results of our analysis of \object{HD 213985} orbital and short-term variability. The proposed mechanism for the  \object{HD 213985} behaviour is described in Section~\ref{sec:discussion}. Our conclusions are presented in Section~\ref{sec:conclusion}.

\section{Data}\label{sec:data}

\begin{figure*}
    \centering
    \includegraphics[width=\textwidth]{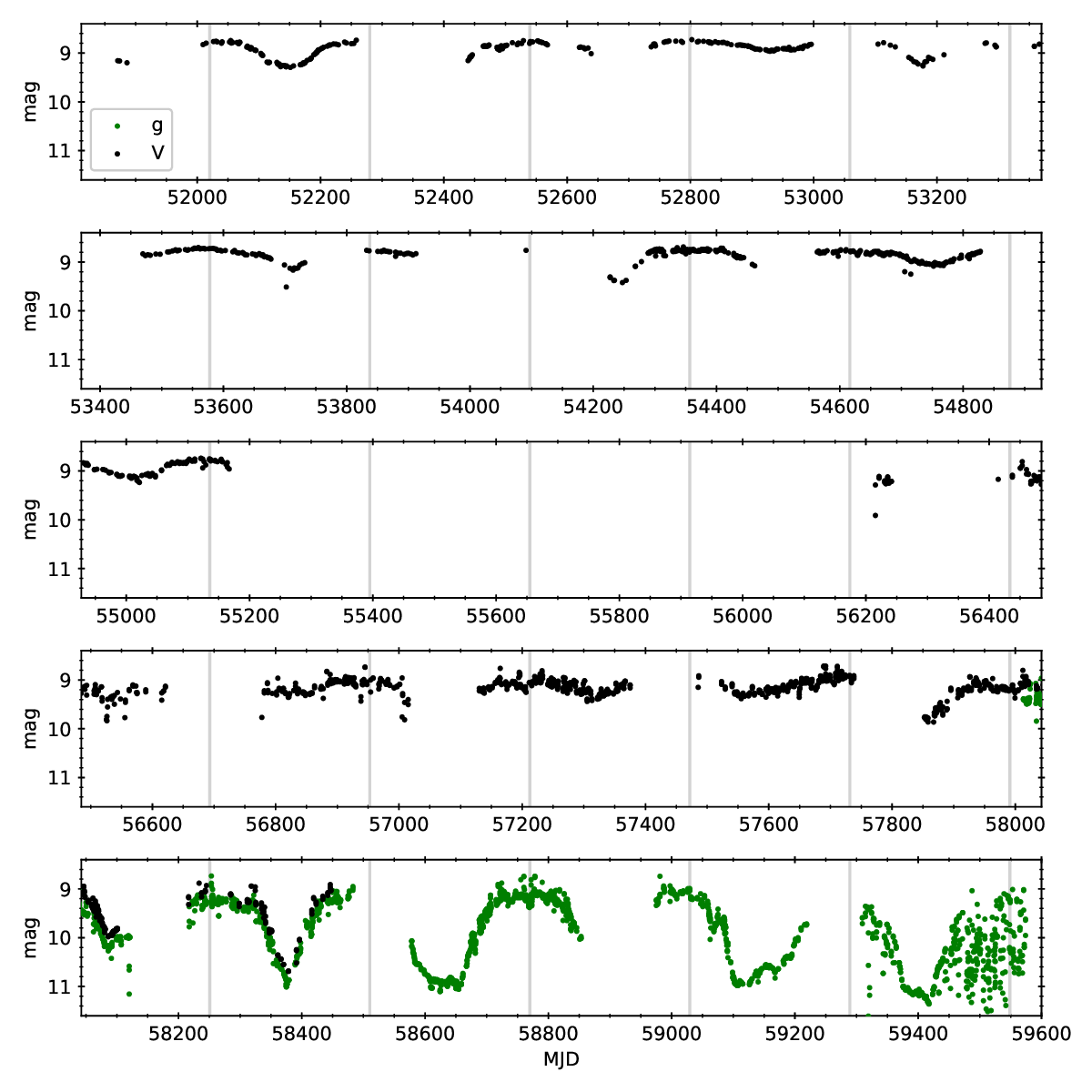}
    \caption{Ground-based light curve of \object{HD 213985}. Black and green points represent observations in the $V$ and $g$ filters, respectively. The grey lines indicate an orbital phase of zero, corresponding to the times of periastron passage \citep{2018A&A...620A..85O}. The data quality declines after MJD = 59300, with increased instrumental scatter. }
    \label{fig:fullLC}
\end{figure*}

\subsection{Photometric observations}
We compiled ground-based photometric observations in $V$ band from the All Sky Automated Survey (ASAS; \citealt{1997AcA....47..467P,2005AcA....55...97P}), as well as in $g$ and $V$ filters from the All-Sky Automated Survey for Supernovae (ASAS-SN; \citealt{2014ApJ...788...48S,2017PASP..129j4502K}). The ASAS data was partially analysed by \citet{2007MNRAS.375.1338K}, where the authors have found small changes in the orbital variability. The quality of ASAS-SN $g$-band observations decreases after MJD=59300 and a large instrumental scatter starts to appear, likely due to the star being close to the limiting magnitude of the telescopes. Since the instrumental scatter seems to increase with time, we have limited our analysis to data before MJD=59600.  The studied light curve spans from 19 August 2001 to 20 January 2020. The full light curve is presented in Fig.~\ref{fig:fullLC} and the phase plots are presented in Fig.~\ref{fig:full_phased}.

\begin{figure*}[h!]
    \centering
    \includegraphics[width=\textwidth]{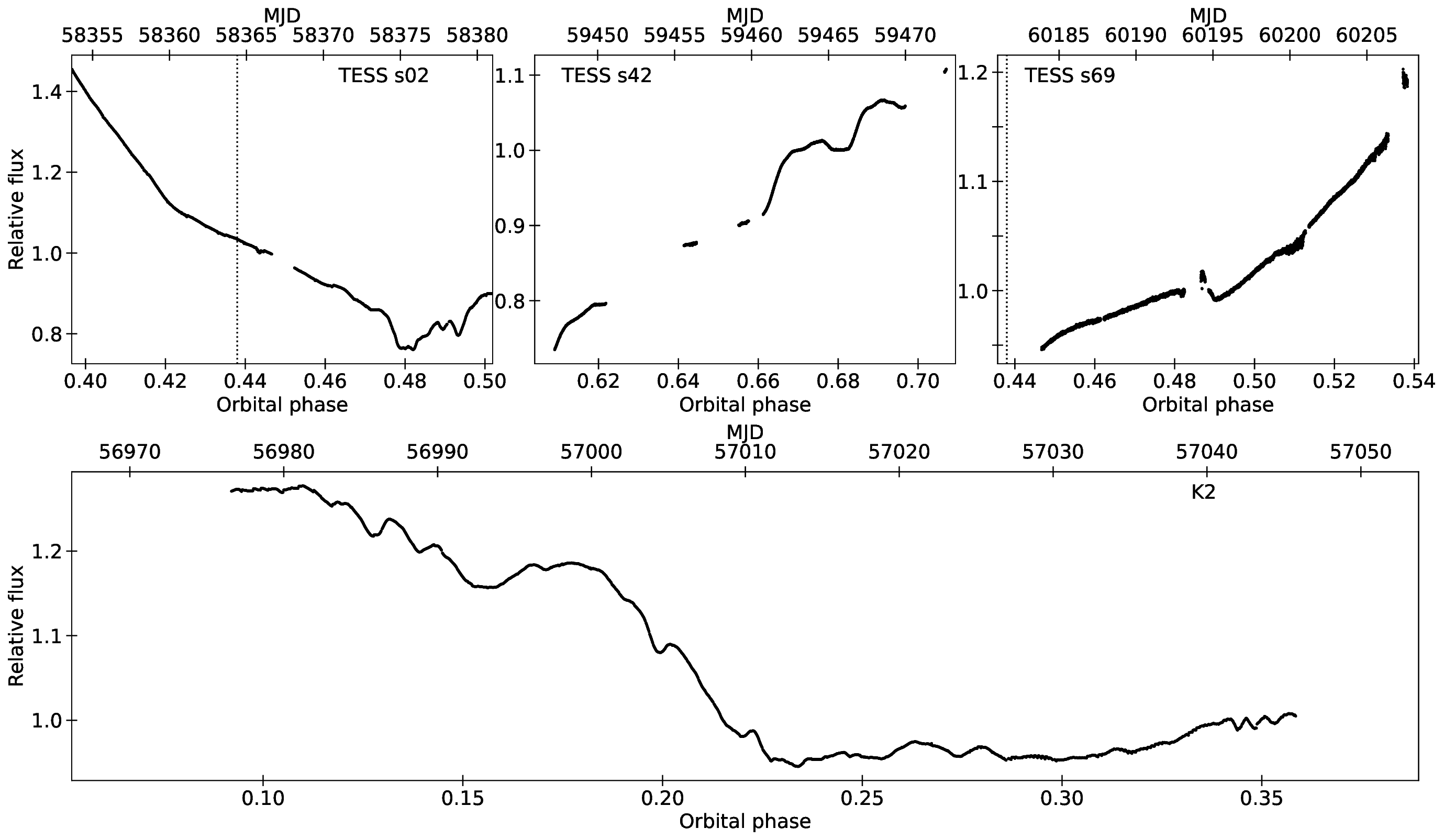}
    \caption{Observations from \textit{K2} and \textit{TESS} sectors 2, 42, and 69. The data were plotted against the \object{HD 213985} orbital ephemeris of \citet{2018A&A...620A..85O}, where phase zero corresponds to the periastron passage and the dotted lines mark the phase of inferior conjunction (post-AGB star is closest to the observer). The fluxes were normalized to the mean flux in each dataset. The decrease in brightness at the orbital phase of 0.49 and a larger scatter of data at the time of inferior conjunction in \textit{TESS} sector 69 data are likely instrumental artefacts.}
    \label{fig:spacedata}
\end{figure*}

We include in our analysis space-based observations from the \textit{K2} mission \citep{2014PASP..126..398H}  covering 15 November 2014 to 23 January 2015, and the \textit{Transiting Exoplanet Survey Satellite} (\textit{TESS}; \citealt{2015JATIS...1a4003R})  data from sectors 2, 42, and 69 (September 2018, 2021, and 2023, respectively)\footnote{\href{https://heasarc.gsfc.nasa.gov/docs/tess/sector.html}{https://heasarc.gsfc.nasa.gov/docs/tess/sector.html}}. The \textit{TESS} light curves have been extracted using the \texttt{Lightkurve} package \citep{2018ascl.soft12013L}. The \textit{K2} light curve is taken directly from the mission data products. In the case of \textit{TESS} data, we have extracted the photometry from the full-frame image cut-outs \citep{2019ascl.soft05007B}.  We have employed simple aperture photometry (SAP) flux  estimates using data from both satellites.  The \textit{K2} and \textit{TESS} light curves are presented in Fig.~\ref{fig:spacedata}.

We also analyze infrared light curves from the Wide-field Infrared Survey Explorer (\textit{WISE}; \citealt{2010AJ....140.1868W,2011ApJ...731...53M}) observations   spanning from 24 May 2010 to 5 November 2023.  The \textit{WISE} observations in $W1$ and $W2$ filters have been retrieved using methods described by \citet{2020MNRAS.493.2271H}\footnote{\href{https://github.com/HC-Hwang/wise_light_curves}{https://github.com/HC-Hwang/wise\_light\_curves}}. The data is binned with a 3~day bin size. In each bin, a weighted average and its error are calculated using individual measurement errors.

\subsection{Spectroscopic observations}
We make use of observations with the HERMES echelle spectograph (R$\sim$85\,000; \citealt{2011A&A...526A..69R}) mounted on a 1.2m Mercator telescope at Roque de los Muchachos  observatory. \object{HD 213985} has been monitored  from  5 August 2009 to 3 January 2024  as part of a post-AGB binaries monitoring programme  \citep{2010MmSAI..81.1022V}. The have already been partially employed for modelling the jets in \object{HD 213985} \citep{2022A&A...666A..40B} and for determining the spectroscopic orbit \citep{2018A&A...620A..85O}.

\section{Results}\label{sec:results}
At the beginning of the observed time period, the orbital variability of \object{HD 213985} was symmetrical with the minimum located roughly at the orbital phase $\sim$0.5, but with a marginally variable amplitude (Fig.~\ref{fig:full_phased}). A deviation from symmetry has started to appear at MJD$\simeq$57300, where the decrease in brightness was larger at phases shortly after the periastron passage  (phase~0). However, the amplitude of the orbital variability remained relatively low ($\Delta$$V$$\lesssim$0.5~mag). This behaviour has already been noted by \citet{2007MNRAS.375.1338K}.

\begin{figure*}[h]
    \centering
    \includegraphics[width=\columnwidth]{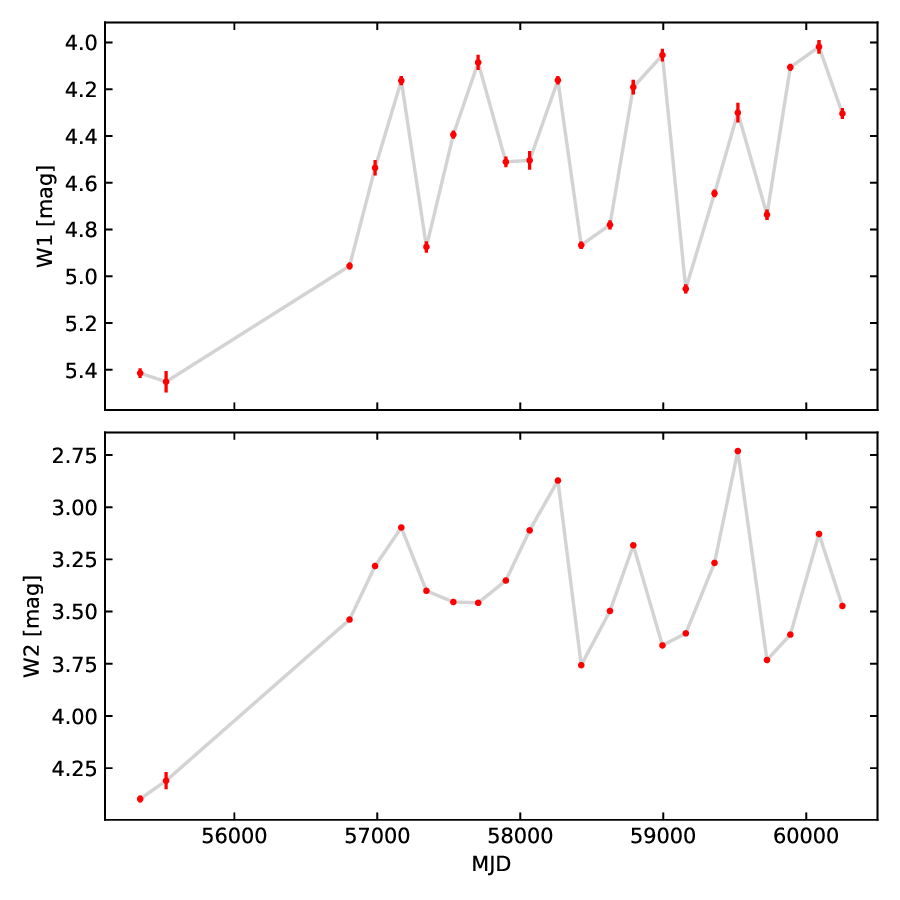}
    \includegraphics[width=\columnwidth]{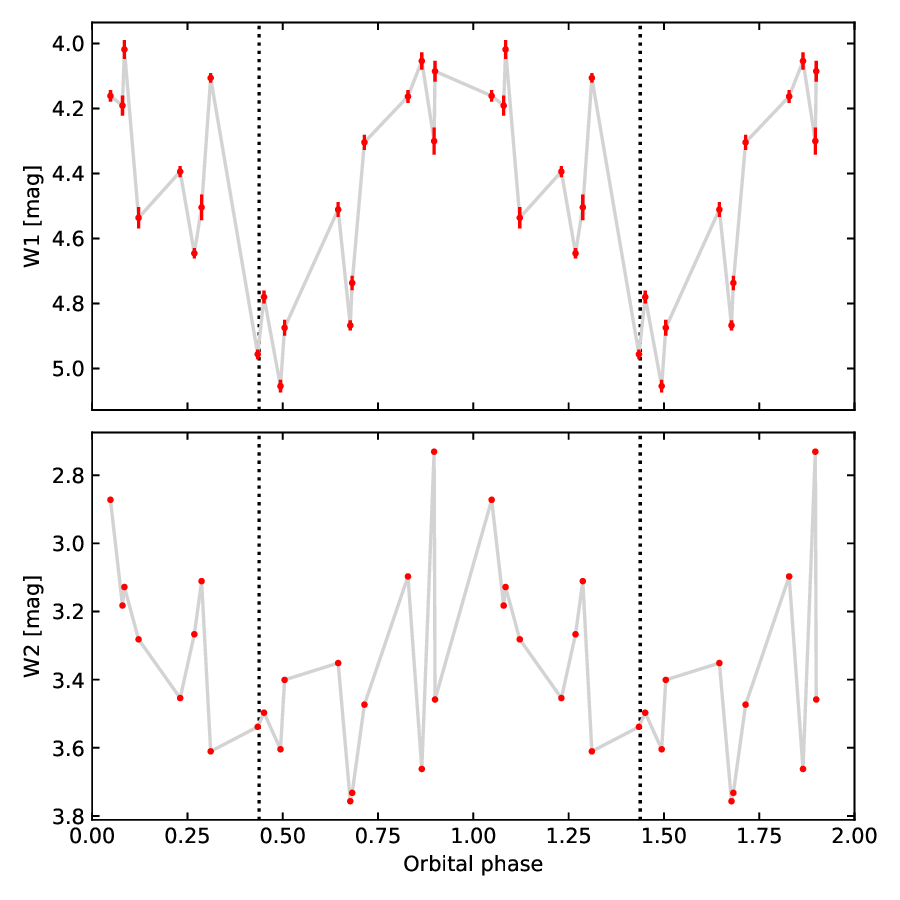}
    \caption{\textit{WISE} photometric measurements (red points). Left panels: Full light curve. Right panels: Phase plot of \textit{WISE} observations made after MJD~56000 using the orbital ephemeris of \citet{2018A&A...620A..85O}, where the dotted lines mark the phase of inferior conjunction (post-AGB star closest to the observer). }
    \label{fig:wise}
\end{figure*}

Since its appearance, the asymmetry in the orbital variability has been present throughout the remaining observing period. The amplitude of variability has started increasing significantly from MJD$\simeq$57850  and has remained at $\Delta$$V$$\simeq$2~mag since MJD$\simeq$58300 (Fig.~\ref{fig:fullLC}). Moreover, the shape of the variability has become more irregular and since MJD$\simeq$58600 a quasi-plateau at the minimum has been present. Despite the emergence of the somewhat irregular variability, if the orbital variability was asymmetrical,  the system is always fainter shortly after the periastron passage (Fig.~\ref{fig:full_phased}). The change in the shape of the orbital variability has been associated with a change in the maximum brightness of \object{HD 213985}. Namely, maximum brightness during an orbital cycle has appeared to decrease from $V$=8.8~mag at MJD$\simeq$52050 to $V$=9.1~mag at MJD$\simeq$58300.

The space-based photometry reveals that \object{HD 213985} experienced episodic small-amplitude short-term fluctuations on timescales of days (Fig.~\ref{fig:spacedata}). This variability was present all throughout the \textit{K2} observing period. On the other hand, during \textit{TESS} sector~2 the star has initially shown a smooth and relatively fast decrease in brightness until orbital phase $\sim$0.42, when the fading has slowed down almost momentarily, but remained smooth afterwards. This continued until orbital phase $\sim$0.47, when the star has faded by 10\% in only 1.05 days. This sudden decrease in brightness was accompanied by the appearance of short-term fluctuations previously seen in \textit{K2}. After six days, at the orbital phase $\sim$0.50, the brightness of \object{HD 213985} has increased to a level slightly larger than the one observed just before the sudden fading. Moreover, the short-term fluctuations seem to cease, although the star was not observed for long enough time to assert that with certainty.

During \textit{TESS} sector~42 observations, \object{HD 213985} brightness has been increasing. Interestingly, short-term fluctuations have appeared throughout the sector, occurring on one of the longer timescales. It is worth noting that the \textit{TESS}  sector~42 dataset is the only time the short-term fluctuations were observed in the second half of the orbital cycle. Moreover, the longer-timescale fluctuations in $K2$ data have appeared to occur closer to the periastron passage. Hence, there is a possibility of a correlation between the timescale of the short-term fluctuations and the orbital phase. However, there is not enough data to claim that with any confidence.

In the overlapping phases of the \textit{TESS} observations from sectors 2 and 69, the system has behaved radically differently. In particular, during \textit{TESS} sector~69 observations, \object{HD 213985} has shown a smooth increase in brightness with no signs of short-term fluctuations, while the opposite is true for sector 2. Notably, it appears that no changes in the smooth increase in brightness have been present at the orbital phase $\sim$0.47 at which a sudden fading has been observed during sector~2.  We note that in \textit{TESS} sector~69 data there is a large scatter and a fading at orbital phase $\sim$0.49, but it is likely due to instrumental artefacts.

\subsection{Infrared light curve}\label{sec:infr}

The brightness of \object{HD 213985} has increased in  $W1$ and $W2$ filters by $\sim$1~mag between MJD$\sim$55000 and $\sim$57000 (Fig.~\ref{fig:wise}). Interestingly, the increase in infrared brightness has occurred roughly at the same time as the appearance of asymmetry in the orbital variability observed in the optical range, and decrease in the maximum brightness in the $V$ filter. After the increase in the infrared brightness, the object has shown variability with amplitude of $\sim$0.5~mag, although the variability in the $W1$ and $W2$ filters appears only partially correlated. 

The low cadence of the infrared photometry makes it impossible to study cycle-to-cycle variations of the orbital variability, and whether it is  similar to what is observed in the optical range. However, after combining all the \textit{WISE} observations after MJD~56000 the overall shape of the phased light curves has become visible (Fig.~\ref{fig:wise}). In particular, the orbital variability in $W1$ filter is readily present, while it is not present or at least less pronounced in the $W2$ band. While both these filters are dominated by the light of the circumbinary disc \citep{2022A&A...658A..36K}, the $W1$ band captures a larger fraction of the post-AGB starlight compared to $W2$. Hence, this is consistent with the orbital variability  being due to the obscuration of the post-AGB star.

\begin{figure*}[h]
    \centering
    \includegraphics[width=\columnwidth]{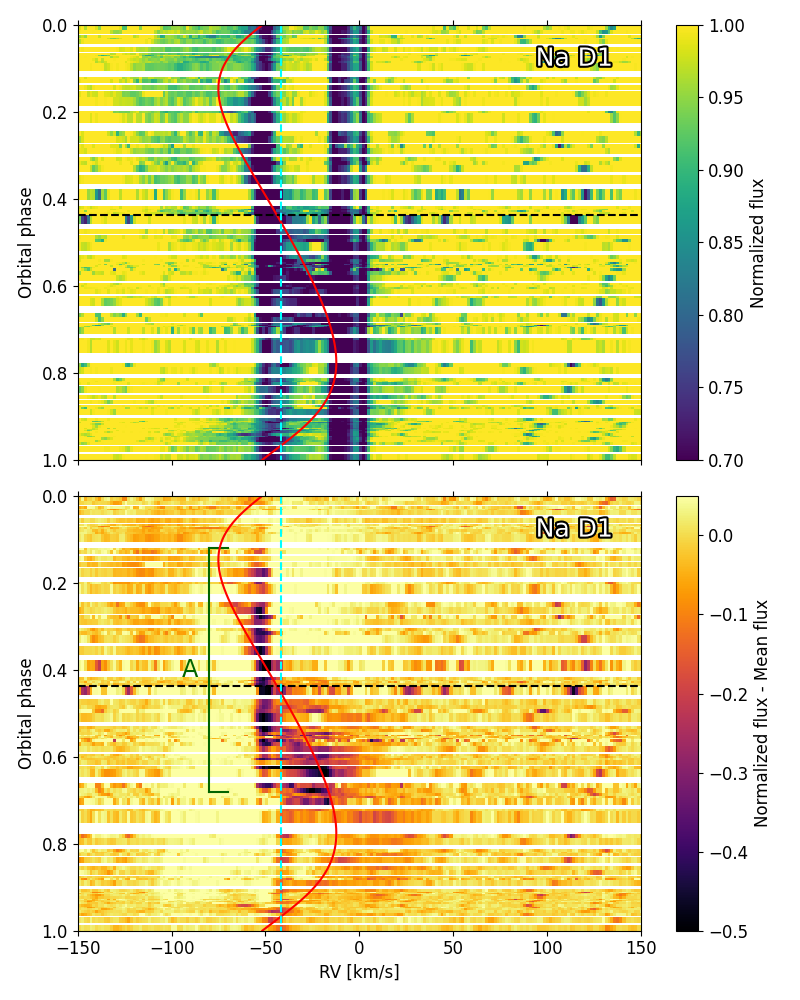}
    \includegraphics[width=\columnwidth]{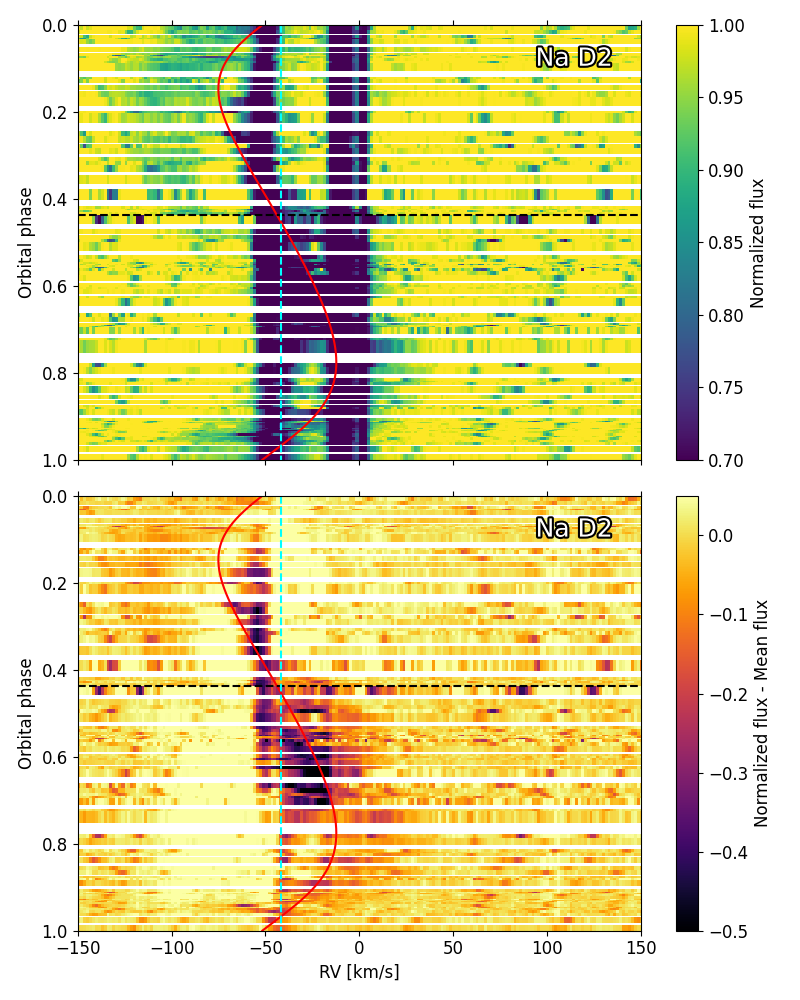}
    \caption{Dynamical spectra of Na~D lines. Top panels show the two Na~D lines. Bottom panels show the same spectra after subtracting the average spectrum at orbital phases $\pm$0.08 of the periastron passage. The red line represents the post-AGB component radial velocity curve according to the spectroscopic orbit of \citet{2018A&A...620A..85O}. The systemic radial velocity is marked with a dashed blue line, and the time of inferior conjunction is marked with a dashed black line. Orbital phase 0.0 corresponds to the periastron passage. The green line marks a time of enhanced absorption from the circumbinary disc component (hereafter region~A).}
    \label{fig:NaD}
\end{figure*}

\subsection{Na~D lines}

We have extracted \ion{Na}{I}~D lines from the \object{HD 213985} spectra and have tracked their variability with the orbital cycle. As expected, the Na~D1 ($\lambda$5895) and D2 ($\lambda$5889) lines behaviour is identical (Fig.~\ref{fig:NaD}). Since D1 component has twice lower oscillator strength than D2, this similarity implies optically thin absorption. Their profiles consist mainly of a few interstellar components at radial velocities between --20 and 6~km\,s$^{-1}$, a weak absorption line originating from the post-AGB star and following its radial velocity that is mainly visible at $-100$~km\,s$^{-1}$ around orbital phase $\sim$0.2 and at +30~km\,s$^{-1}$ half an orbit later, as well as a strong circumbinary component roughly at the systemic velocity of the binary.  The circumbinary component has been observed at the radial velocity of approximately -55~km\,s$^{-1}$. The circumbinary component is blueshifted by $\sim$10~km\,s$^{-1}$ relative to the systemic velocity, similarly to what is observed in a post-AGB binary \object{BD+46$\degr$442} \citep{2012A&A...542A..27G}. Moreover,  between the orbital phases $\sim$0.5 and 0.7, an additional broad absorption profile has become visible.

To better visualise the variability of Na~D lines, we have subtracted the average spectrum taken at orbital phases $\pm$0.08 of the periastron passage, that is,  when the absorption of various Na~D components was the weakest. As a result, the interstellar components have been removed, and the variability of the circumbinary component has become clearly visible (Fig.~\ref{fig:NaD} bottom). In particular, the strength of the circumbinary component has been strongest around the phases of inferior conjunction, when the post-AGB star is closest to the observer (region A in Fig~\ref{fig:NaD}). The transient absorption profile present between orbital phases $\sim$0.5 and 0.7 appears at the radial velocity of the post-AGB star or slightly blueshifted. Interestingly, there are no strong absorption components from the jet, seen previously in the H$\alpha$ profile at times near the superior conjunction, at orbital phases $\sim$0.97 \citep{2022A&A...666A..40B}.

\section{Discussion }\label{sec:discussion}

The optical and infrared photometric orbital variability of \object{HD 213985} is consistent with variable obscuration by the inner edge of the circumbinary disc, similarly to what is seen in RVb-type RV~Tau stars \citep{2017A&A...608A..99K,2019A&A...628A..40M}.  This interpretation is supported by the variability of the circumbinary component of the Na~D lines, observed at a radial velocity of approximately -55~km\,s$^{-1}$ (Fig~\ref{fig:NaD}). The Na~D absorption is strongest near the time of inferior conjunction, when the post-AGB star is closest to the observer (region A in Fig.~\ref{fig:NaD}). As the equivalent width of Na~D absorption correlates with dust absorption along the line of sight \citep{1997A&A...318..269M,10.1111/j.1365-2966.2012.21796.x}, this confirms that the post-AGB star is obscured by the inner edge of the circumbinary disc during inferior conjunction.

The interpretation of the cycle-to-cycle variability is more complicated.   \citet{2007MNRAS.375.1338K} interpreted these changes as a sign of circumbinary shell/disc variations on timescales similar to the orbital period. This would be consistent with the appearance of the irregular behaviour of the orbital variability roughly at the time when the infrared brightness increased and maximum optical brightness decreased. Moreover, this would be consistent with a large variability of \object{HD 213985} in the $W2$ filter. It is not obviously connected to the orbital variability, but it rather might reflect changes in the circumbinary disc (Fig.~\ref{fig:wise}). A similar interpretation has been proposed for secular changes in the obscuration degree observed in a RV~Tau star \object{U~Mon} \citep{2021ApJ...909..138V}, where the authors proposed precession of the binary or circumbinary disc inhomogeneity caused by binary-disc interaction \citep[see also][]{2018A&A...616A.153K,2020A&A...642A.234O}. However, the secular changes in \object{U~Mon} have appeared to be periodic or quasi-periodic \citep{2021ApJ...909..138V}, while the changes in the shape of \object{HD 213985} orbital variability seem more erratic. Namely, the observations show that orbital variability has switched seemingly randomly between being symmetrical and having a positive skewness, that is,  with the largest fading being shortly after the periastron passage.  Hence, if the circumbinary disc variations are indeed behind the cycle-to-cycle variability in \object{HD 213985}, clumping or 3D structures in the circumbinary disc seem more likely.  

An alternative scenario involves a potential third component in the system, either interacting with the circumbinary disc or influencing the orbital geometry such that the apparent motion of the post-AGB star now trails behind denser regions of the circumbinary disc. However, the system inclination of $i$=50\fdg1$\pm$0\fdg3 \citep{2022A&A...666A..40B} and the well-fitted radial velocity curve \citep{2018A&A...620A..85O} make the possibility of such a third body remaining undetected unlikely.

Another interpretation of the erratic behaviour of \object{HD 213985} could be proposed  based on the Na~D absorption component seen in orbital phases $\sim$0.5--0.7 (Fig.~\ref{fig:NaD}). This component has been observed at the radial velocity of the post-AGB star or slightly blueshifted, at orbital phases when the star starts to move away from the observer. This is similar to the behaviour of \ion{He}{I}~$\lambda$4471 line in a massive interactive binary \object{RY~Sct}, where it was caused by outflow through L2 Lagrange point \citep{2007ApJ...667..505G}. The same interpretation can be valid for \object{HD 213985}, and the cycle-to-cycle orbital shape variability could be explained by interaction of matter ejected through L2 with the circumbinary disc. The strongest outflow would occur during the periastron passage, but it is possible that the average amount of matter lost trough L2 would vary between orbital cycles. However, the post-AGB star in \object{HD 213985} is not filling its Roche Lobe and the exact physics and strength of an outflow through L2 would need to be understood in order to verify this scenario.

Space-based observations have revealed short-term fluctuations of \object{HD 213985}. Similar variability on timescales of days has been observed in a few other post-AGB objects \citep{2012AstL...38..157A,2014AstL...40..485A,2024A&A...682A.133K}. \citet{2024A&A...682A.133K} speculated that the fast variability in \object{U~Equ} can be interpreted as flickering, a stochastic variability seen in all types of astrophysical objects with accretion \citep[e.g.][]{2015SciA....1E0686S}. However, during \textit{TESS} sector~2 observations, the  amplitude of \object{HD 213985} short-term fluctuations increases when the system fades at the orbital phase $\sim$0.47 (Fig.~\ref{fig:spacedata}). This is opposite to what is expected from flickering \citep[e.g.][]{2005MNRAS.359..345U}. Moreover, the light contribution of the circumcompanion disc is estimated to be very minimal (see e.g. \citealt{2018A&A...616A.153K} where for a post-AGB binary  \object{IRAS08544-4431} it is resolved to be 3.9$\pm$0.7\% in the H-band). Flickering of the circumcompanion disc will therefore not introduce such large amplitude fluctuations as have been observed in \object{HD 213985}. On the other hand, \citet{2012AstL...38..157A} suggested that a similar fast variability in  a post-AGB star \object{IRAS 19336-0400} is due to sporadic changes in the mass outflow rate. This interpretation seems to not fit in the case of \object{HD 213985}, as we presume the outflow would be the strongest at or near the periastron passage, as it would be enhanced by tidal forces. The opposite seems to be observed in \object{HD 213985}, where there is no correlation between the presence of fast variability and the orbital phase. This is best seen by presence of the stochastic variability during \textit{TESS} sector 2 observations, and its absence at the same orbital phases during the sector 69 monitoring (Fig.~\ref{fig:spacedata}).

We propose that the short-term fluctuations of \object{HD 213985} may be caused by small dust structures or clumps in the circumbinary disc.  Their presumed evolution and movement would cause variable line-of-sight reddening component on top of the orbital motion induced global line-of-sight reddening.   Obscuration by irregular dust structures is consistent with the fast variability present during \textit{TESS} sector~2, where the variability appears when the system suddenly faded, and disappeared when the brightness increased.  Similar scenario has been proposed to explain the fast variability of a pre-main sequence binary \object{CoRoT 223992193} \citep{2015MNRAS.454.3472T}. 
Based on the spectroscopic orbit of \citet{2018A&A...620A..85O}, the post AGB star is expected to take approximately five days to pass behind a stationary dust cloud that is significantly smaller than its own radius. This aligns with the timescale of the longer fluctuations, implying the upper limit on the size of the dust structures (Fig.~\ref{fig:spacedata}). The shorter fluctuations, occurring on a timescale of one day, would need to be explained either by changes in the dust cloud geometry or by dust clouds counter rotating relative to the binary orbit. A lower limit on the dust structures size can be estimated assuming that the dust structures are completely opaque. A fading by 10\% suggests that 10\% of post-AGB star surface area would be obscured, implying a lower limit of the dust structure radius of $\sqrt{0.1}$ times that of the post-AGB star. Given the post-AGB star radius of $8.6\pm0.3$~R$_\odot$ \citep{2022A&A...666A..40B}, the expected dust structure radius is in the range 2.6--8.9~R$_\odot$.  Fast multi-colour photometry is required to confirm variable line-of-sight reddening as a source of \object{HD 213985} fast variability.

\section{Conclusions}\label{sec:conclusion}

\object{HD 213985} was known to exhibit relatively small variations in the shape of orbital variability \citep{2007MNRAS.375.1338K}. We have shown that the amplitude of \object{HD 213985} variability increased in the recent years. Moreover, its shape has become more irregular, and the orbital variability has often appeared skewed (Fig.~\ref{fig:fullLC}). Based on the variability of Na~D lines we propose that the variability is caused by eclipse by the inner edge of circumbinary disc that is possibly perturbed by outflow through L2 Lagrange point.

The system experienced short-term fluctuations. This variability seems to appear and disappear seemingly at random. In particular, its presence was not related to the orbital phase. Moreover, it could be detected for $\sim$70 days in \textit{K2} and disappear only six days after its appearance during \textit{TESS} sector 2 (Fig.~\ref{fig:spacedata}). The behaviour of fast variability seems consistent with obscuration by small and irregular dust structures, likely located in the circumbinary disc.

The analysis of \object{HD 213985} photometric observations revealed that its behaviour is more complex than previously thought. High-cadance infrared monitoring of the system covering a few orbital cycles is required in order to confirm whether the changes in the orbital variability are caused by variable circumbinary disc. Moreover, simultaneous high-time-resolution spectroscopic and photometric observation are needed to confirm the nature of the short-term fluctuations.

\begin{acknowledgements}
This work was supported by Polish National Science Center grant Sonatina 2021/40/C/ST9/00186. L.P. is a FNRS research fellow.

Based on observations made with the Mercator Telescope, operated on the island of La Palma by the Flemish Community, at the Spanish Observatorio del Roque de los Muchachos of the Instituto de Astrofísica de Canarias. Based on observations obtained with the HERMES spectrograph, which is supported by the Research Foundation - Flanders (FWO), Belgium, the Research Council of KU Leuven, Belgium, the Fonds National de la Recherche Scientifique (F.R.S.-FNRS), Belgium, the Royal Observatory of Belgium, the Observatoire de Genève, Switzerland and the Thüringer Landessternwarte Tautenburg, Germany.

This research made use of Lightkurve, a Python package for Kepler and TESS data analysis \citep{2018ascl.soft12013L}.
\end{acknowledgements}

\bibpunct{(}{)}{;}{a}{}{,} 
\bibliographystyle{aa} 
\bibliography{references}

\begin{appendix}
\section{Ground-based photometry}

   \begin{figure}
   \centering
   \includegraphics[width=\columnwidth]{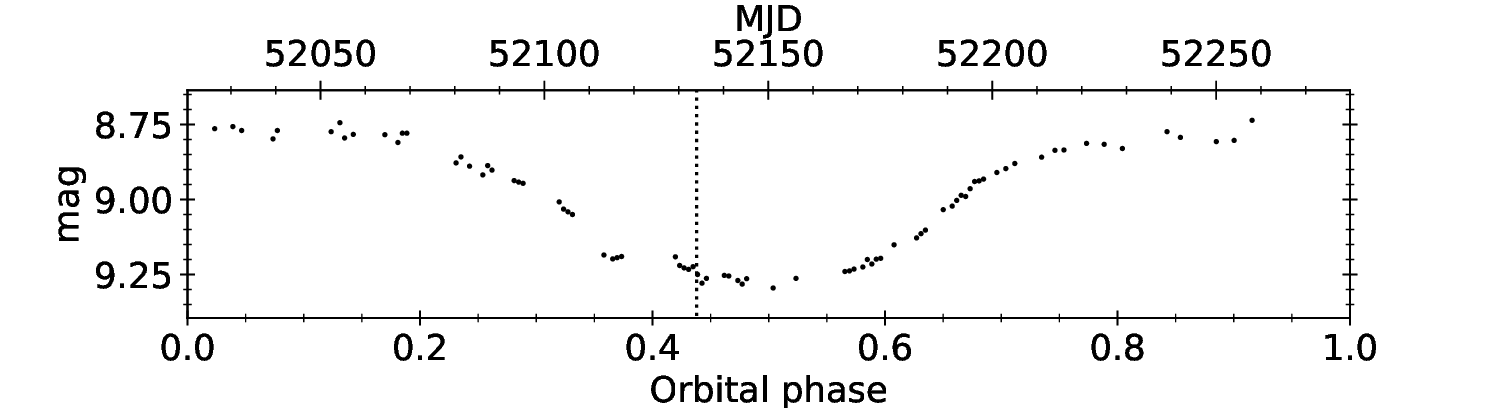}
   \includegraphics[width=\columnwidth]{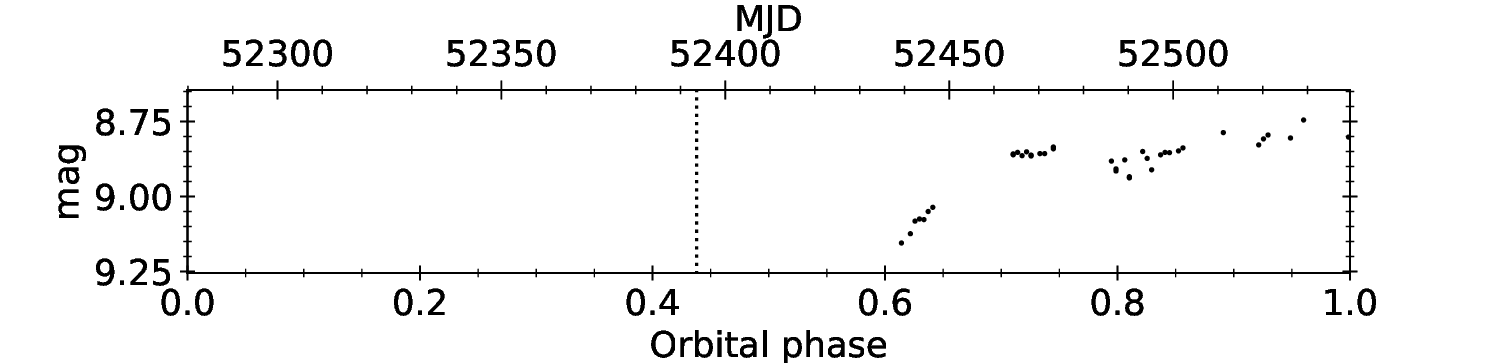}
   \includegraphics[width=\columnwidth]{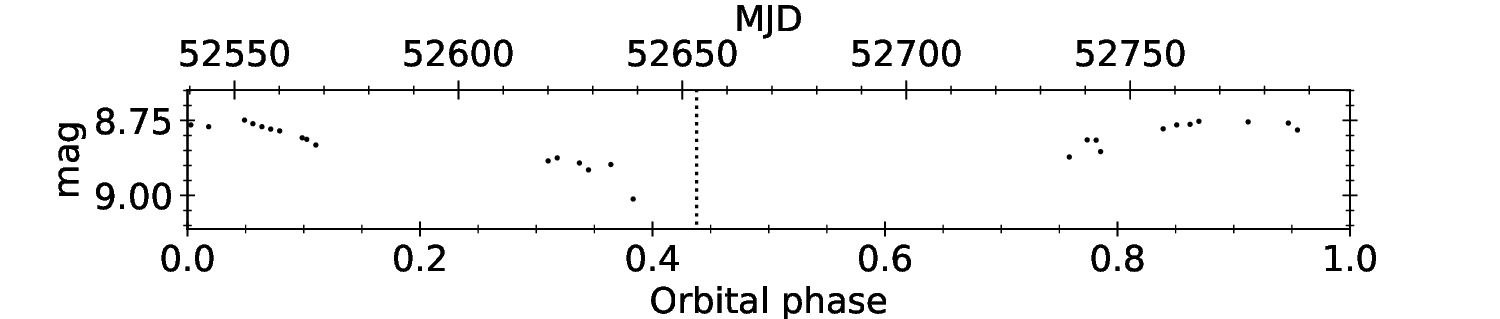}
   \includegraphics[width=\columnwidth]{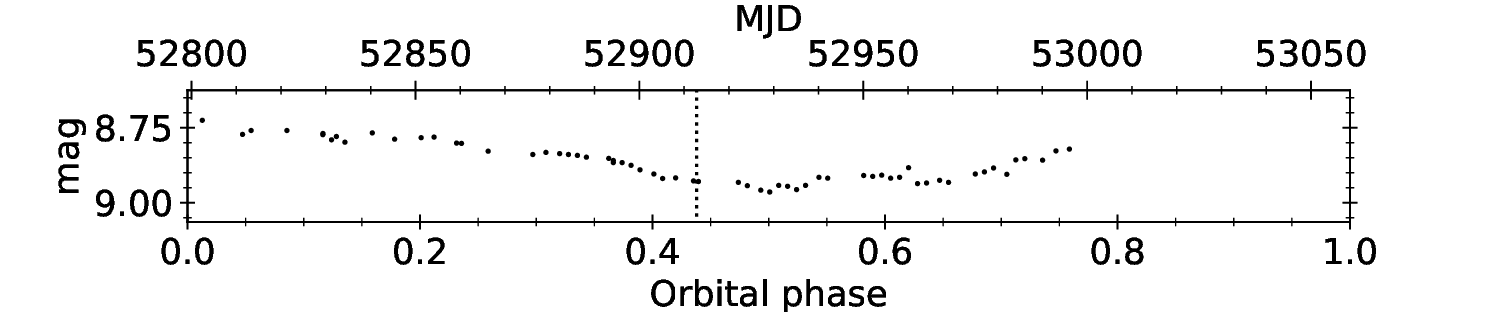}
   \includegraphics[width=\columnwidth]{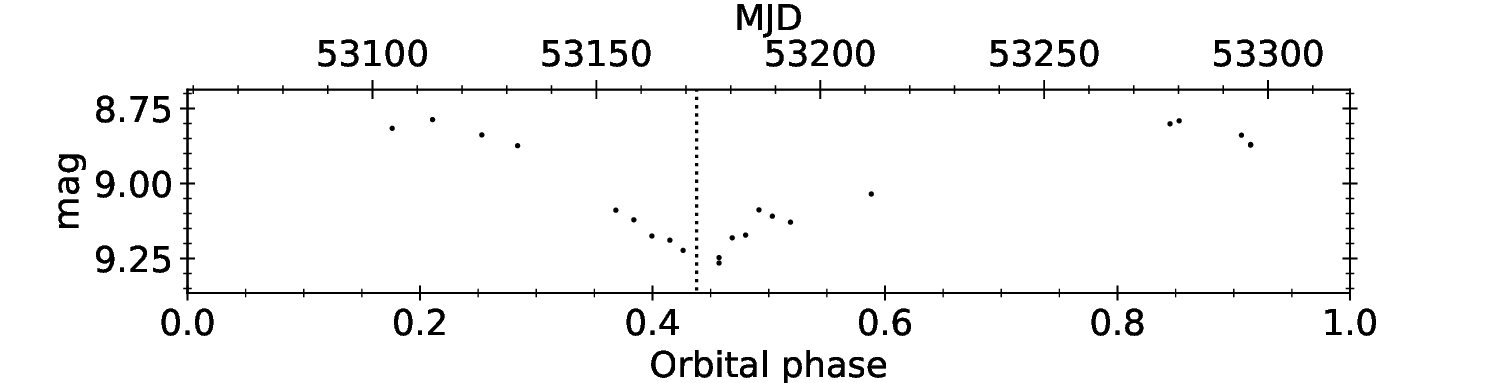}
   \includegraphics[width=\columnwidth]{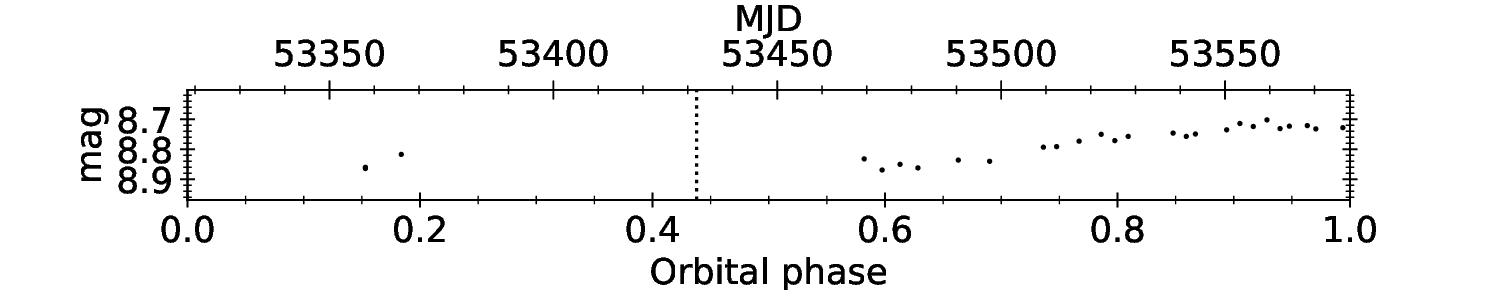}
   \includegraphics[width=\columnwidth]{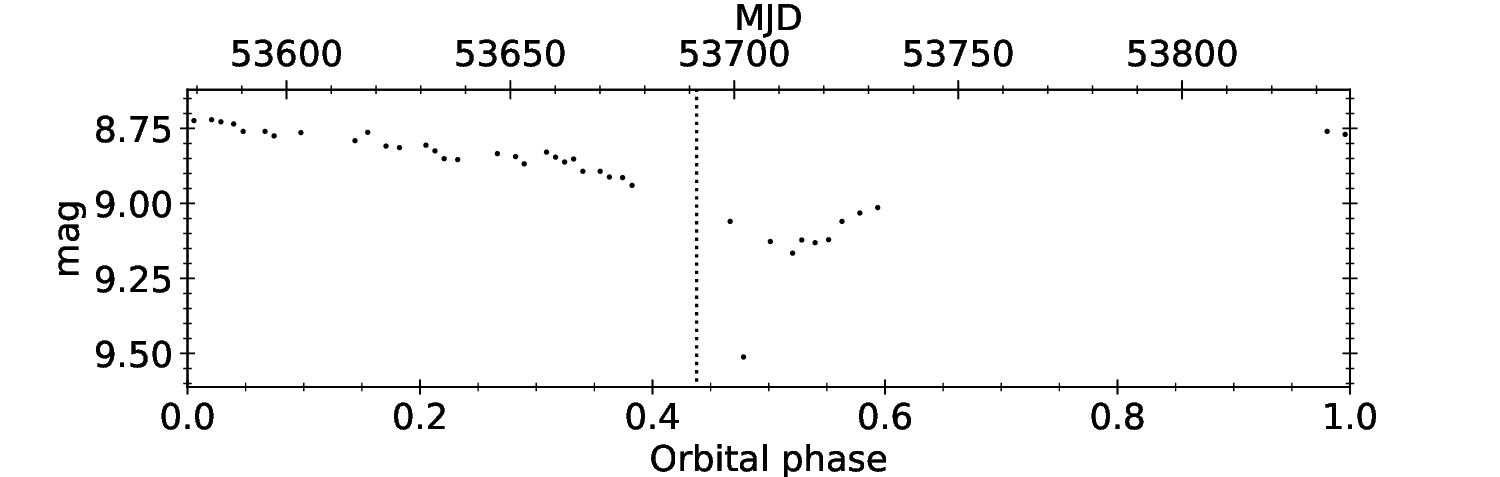}
   \includegraphics[width=\columnwidth]{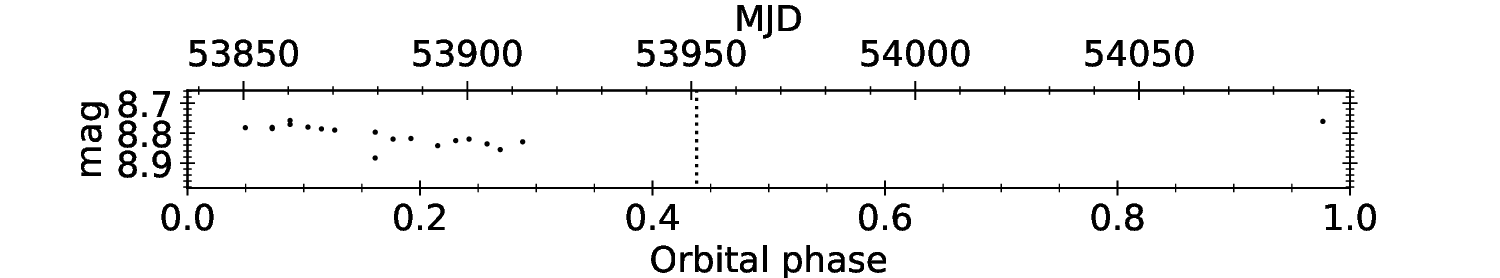}
      \caption{Phase plots of each observed orbital cycle. The data was plotted against the \object{HD 213985} orbital ephemeris of \citet{2018A&A...620A..85O}, where phase zero corresponds to the periastron passage and the dotted lines mark the phase of inferior conjunction (post-AGB star closest to the observer). Black and green points are datapoints in $V$ and $g$ filters, respectively. The grey areas mark the times of \textit{K2} and \textit{TESS} observations.}
         \label{fig:full_phased}
   \end{figure}

   \begin{figure}
   \ContinuedFloat
   \centering
   \includegraphics[width=\columnwidth]{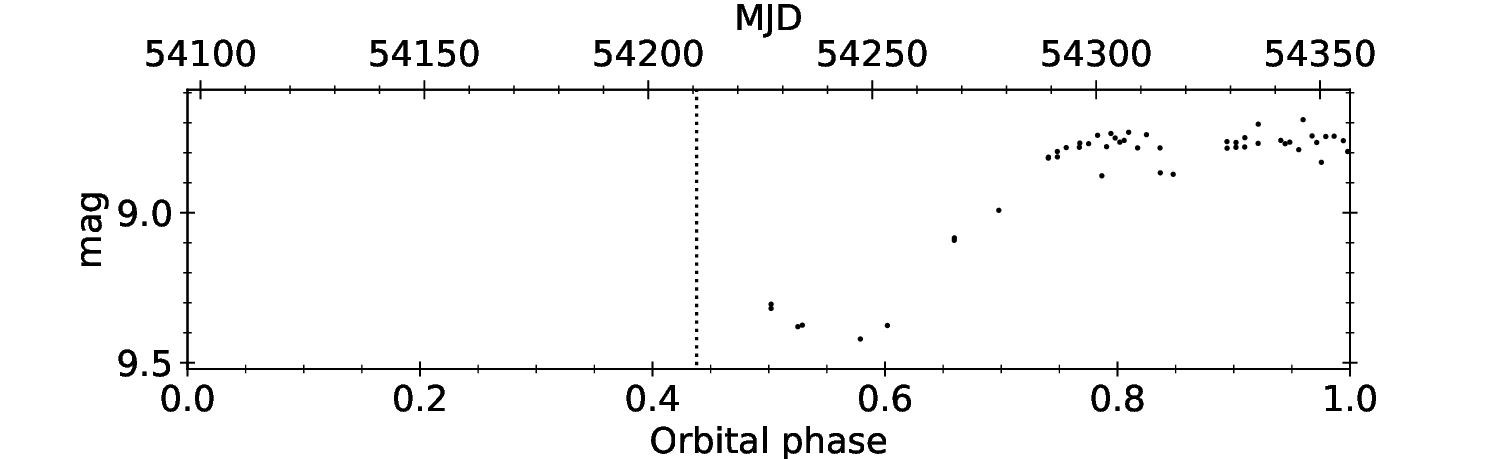}
   \includegraphics[width=\columnwidth]{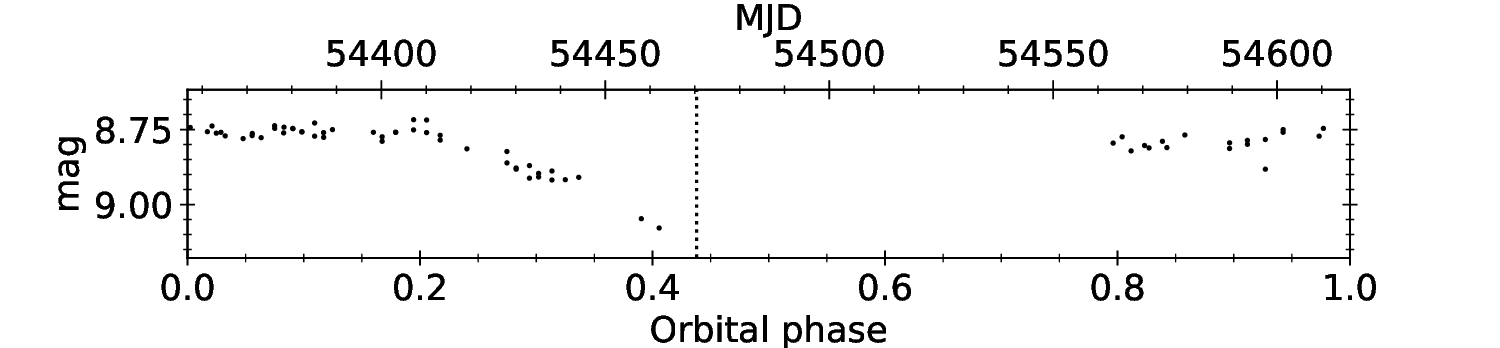}
   \includegraphics[width=\columnwidth]{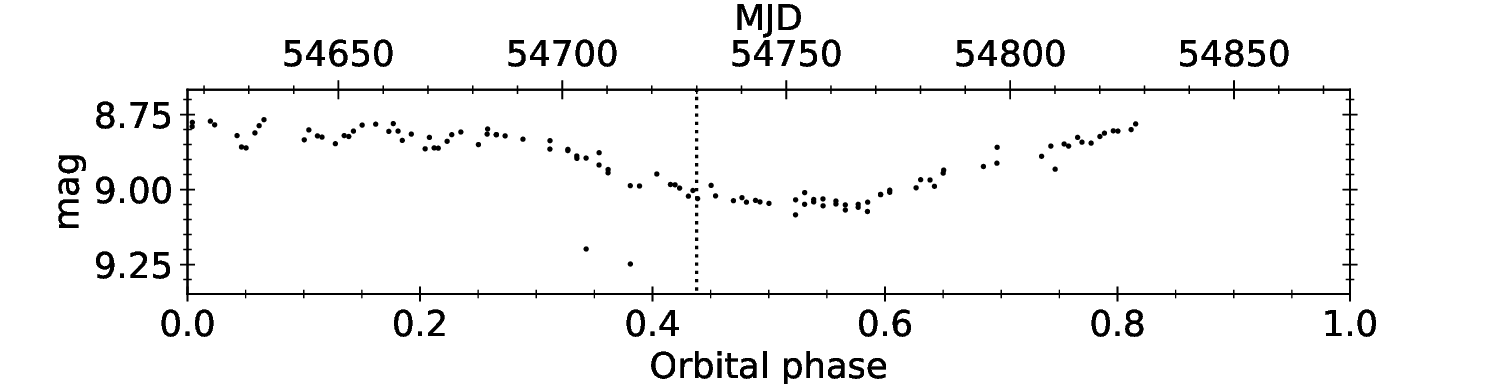}
   \includegraphics[width=\columnwidth]{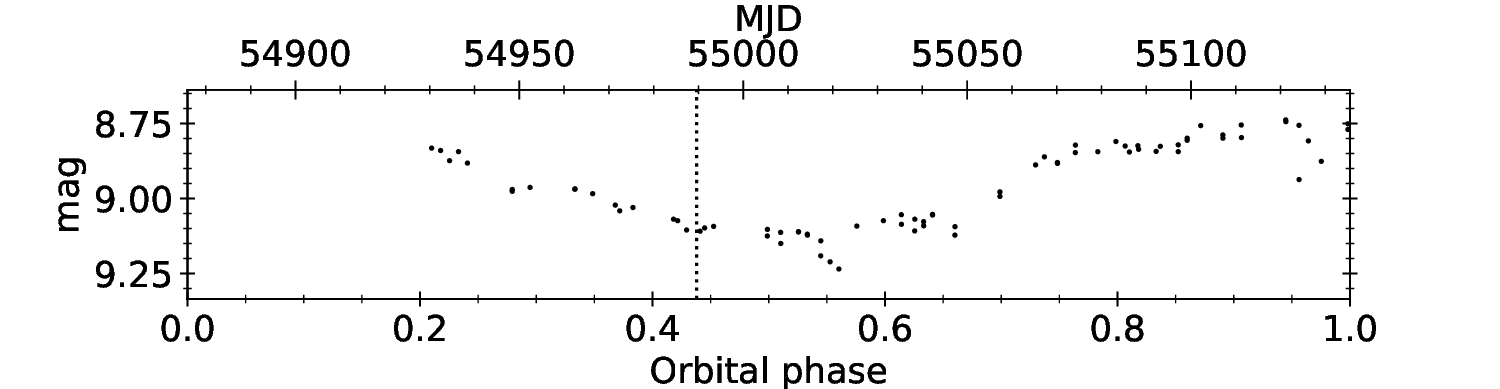}
   \includegraphics[width=\columnwidth]{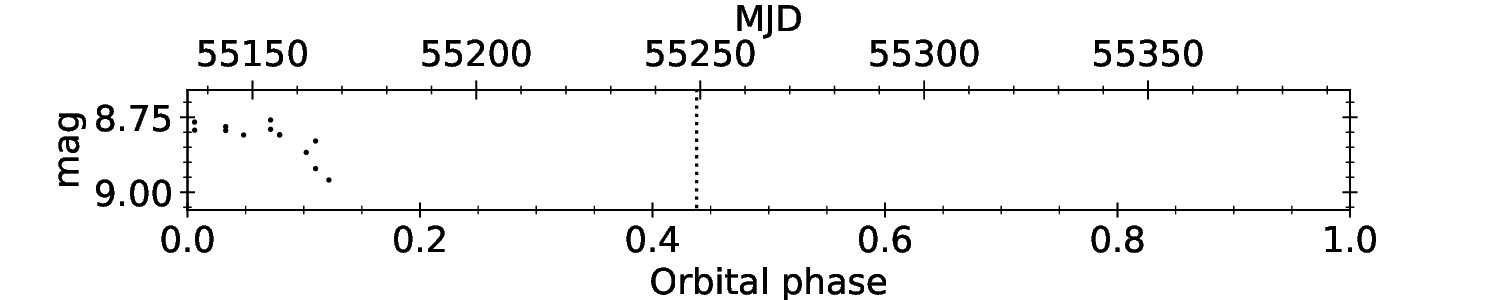}
   \includegraphics[width=\columnwidth]{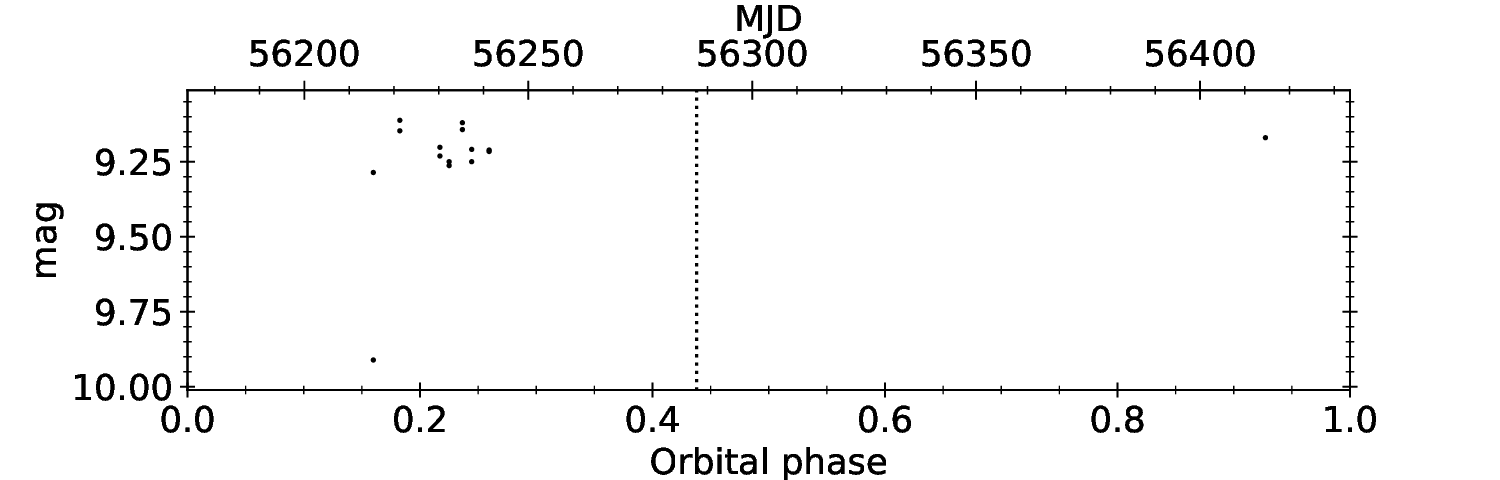}
   \includegraphics[width=\columnwidth]{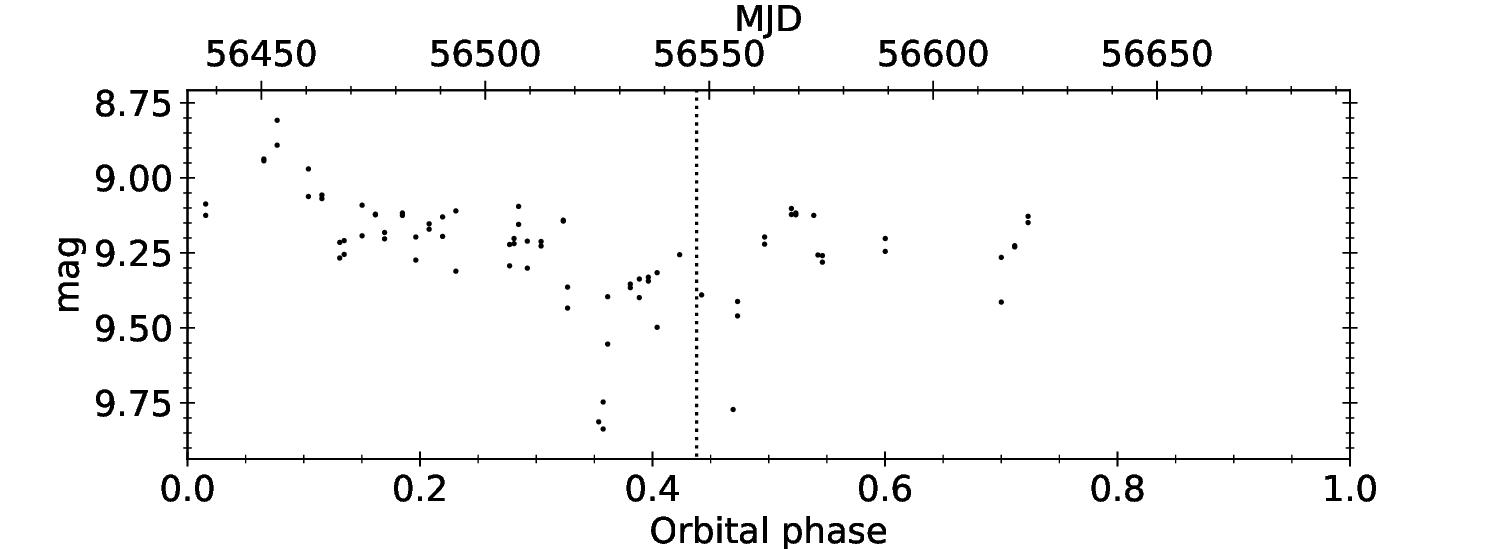}
      \caption{Continued.}
   \end{figure}

   \begin{figure}
   \ContinuedFloat
   \centering
   \includegraphics[width=\columnwidth]{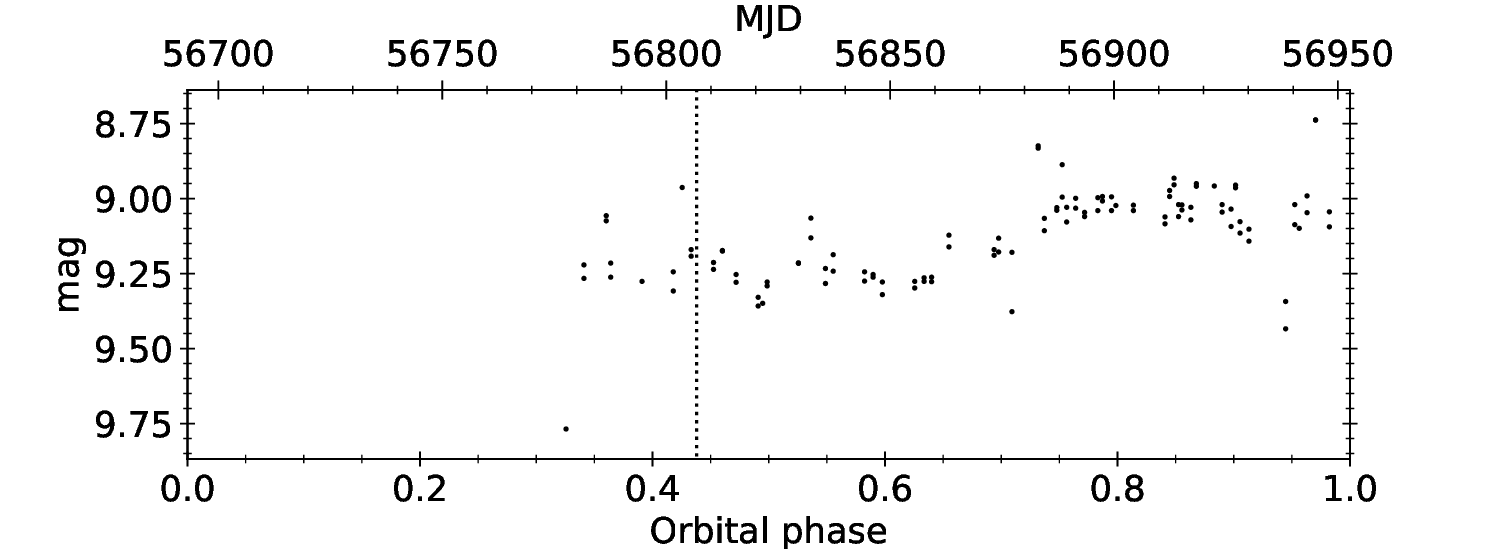}
   \includegraphics[width=\columnwidth]{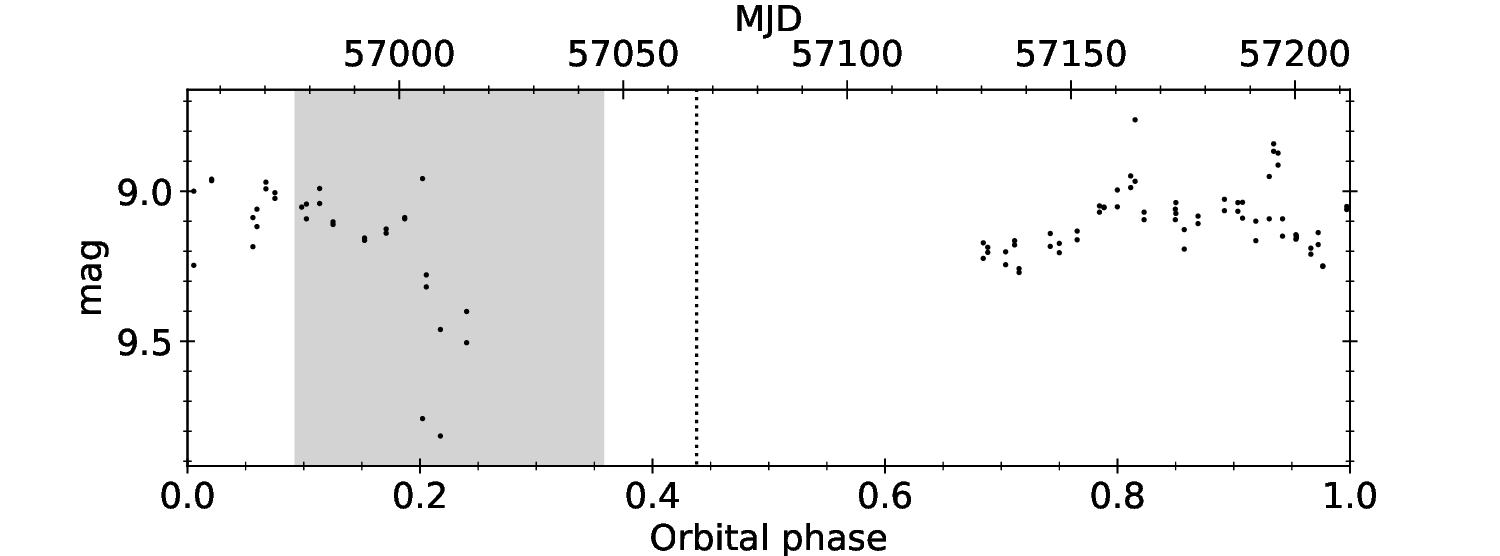}
   \includegraphics[width=\columnwidth]{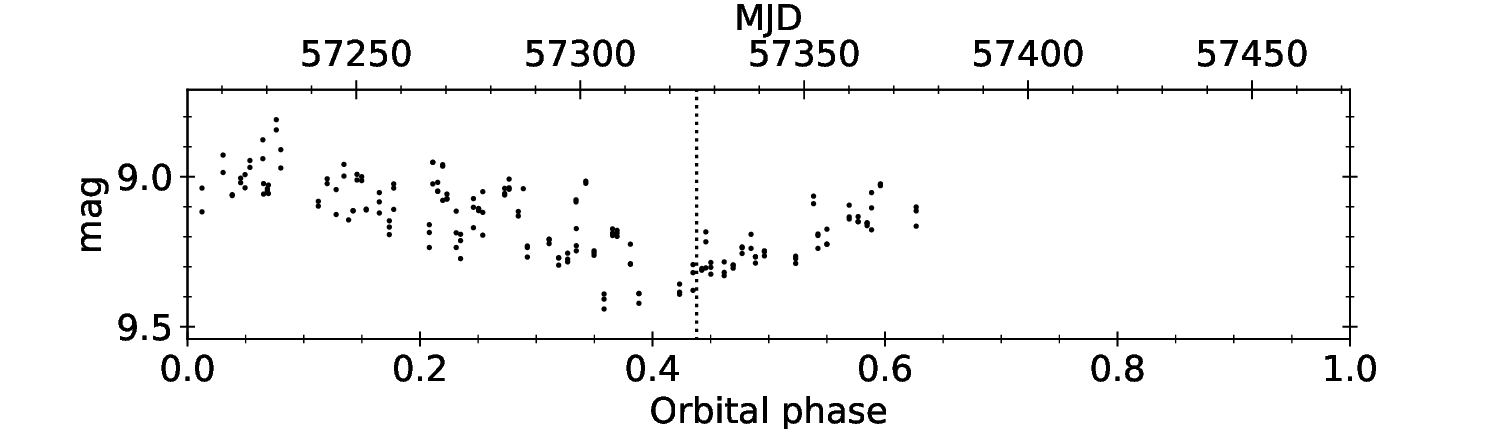}
   \includegraphics[width=\columnwidth]{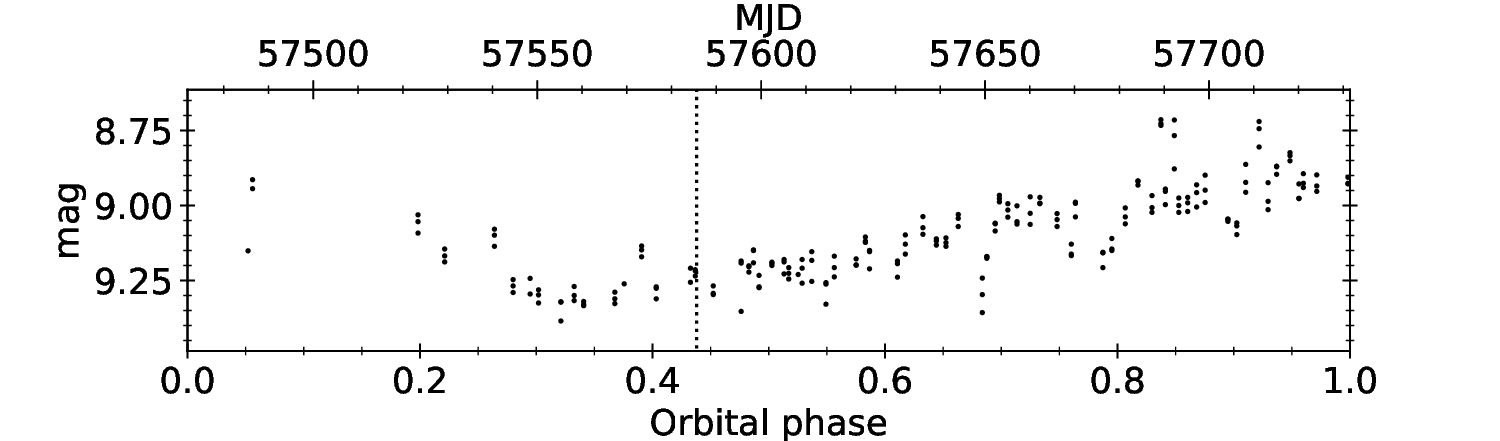}
   \includegraphics[width=\columnwidth]{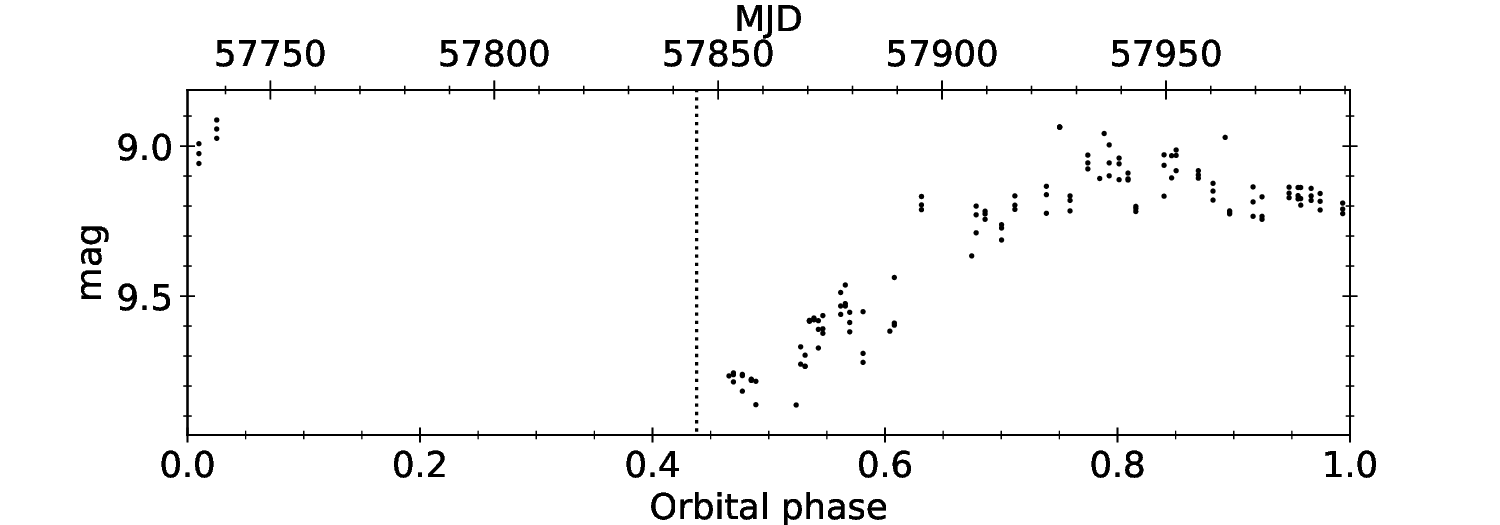}
   \includegraphics[width=\columnwidth]{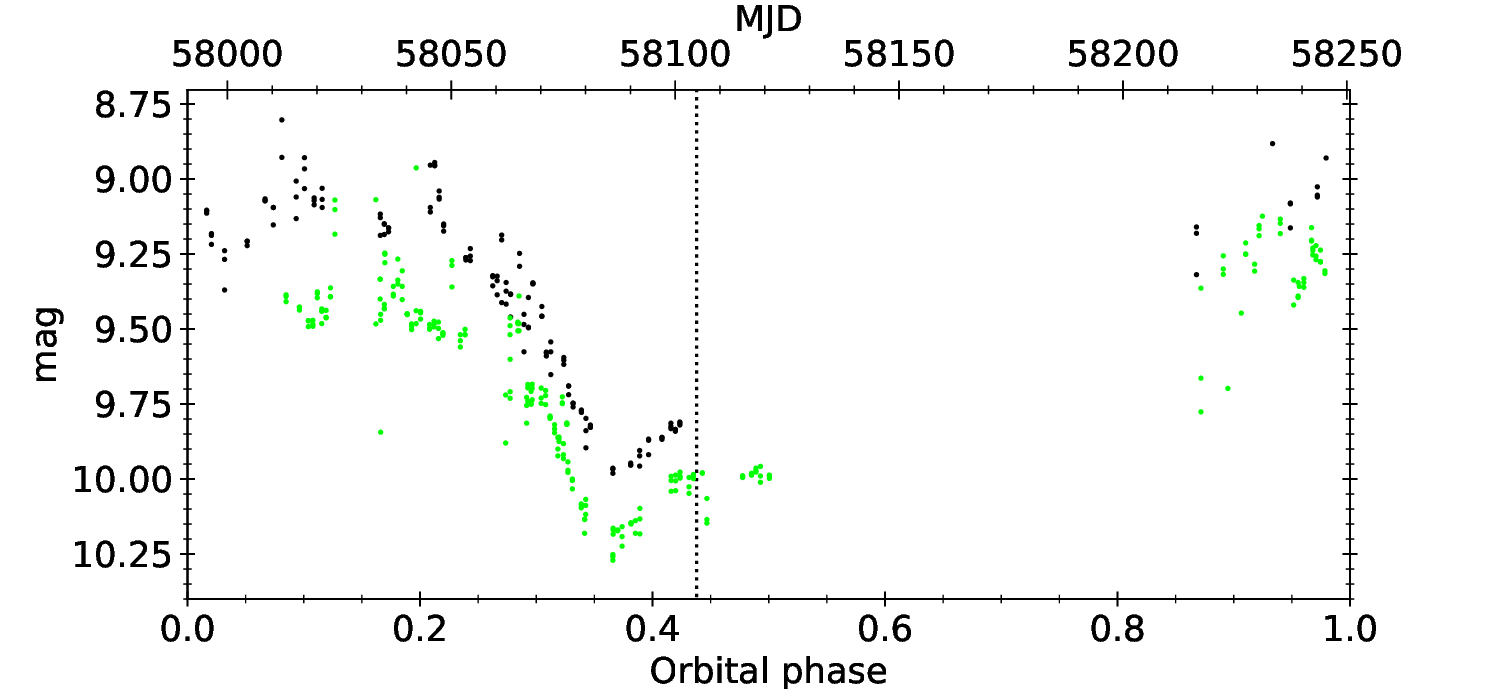}
      \caption{Continued.}
   \end{figure}

   \begin{figure}
   \ContinuedFloat
   \centering
   \includegraphics[width=\columnwidth]{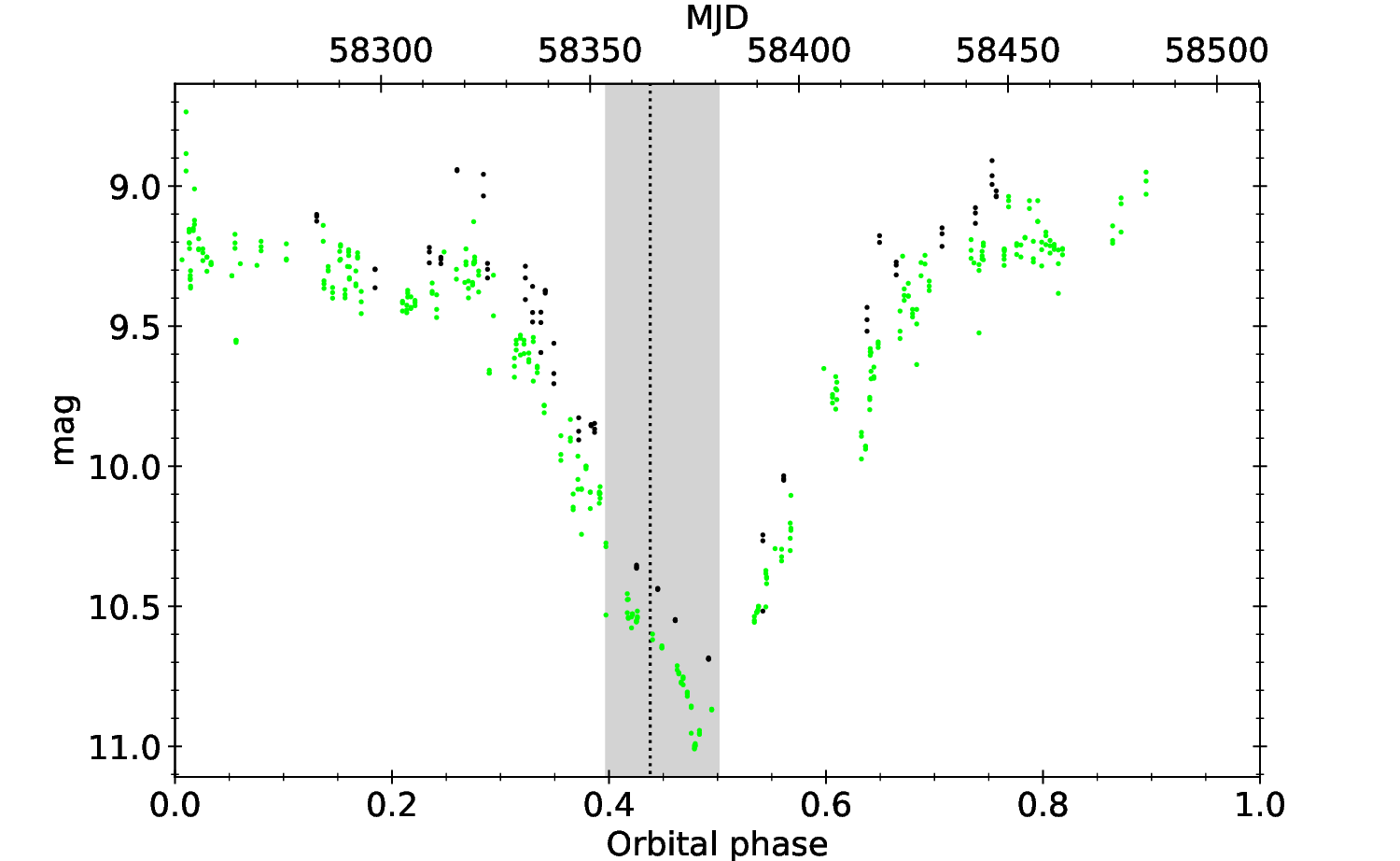}
   \includegraphics[width=\columnwidth]{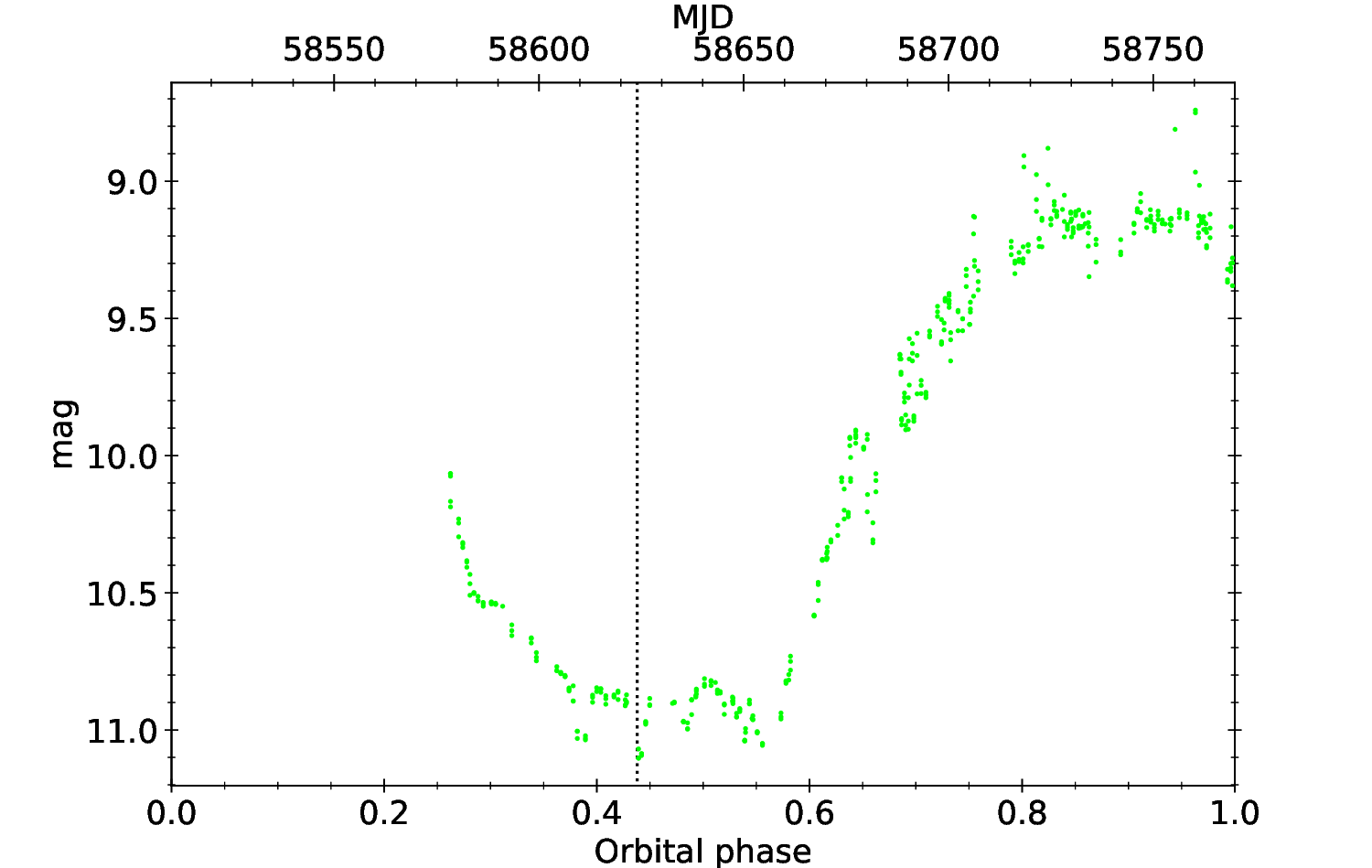}
   \includegraphics[width=\columnwidth]{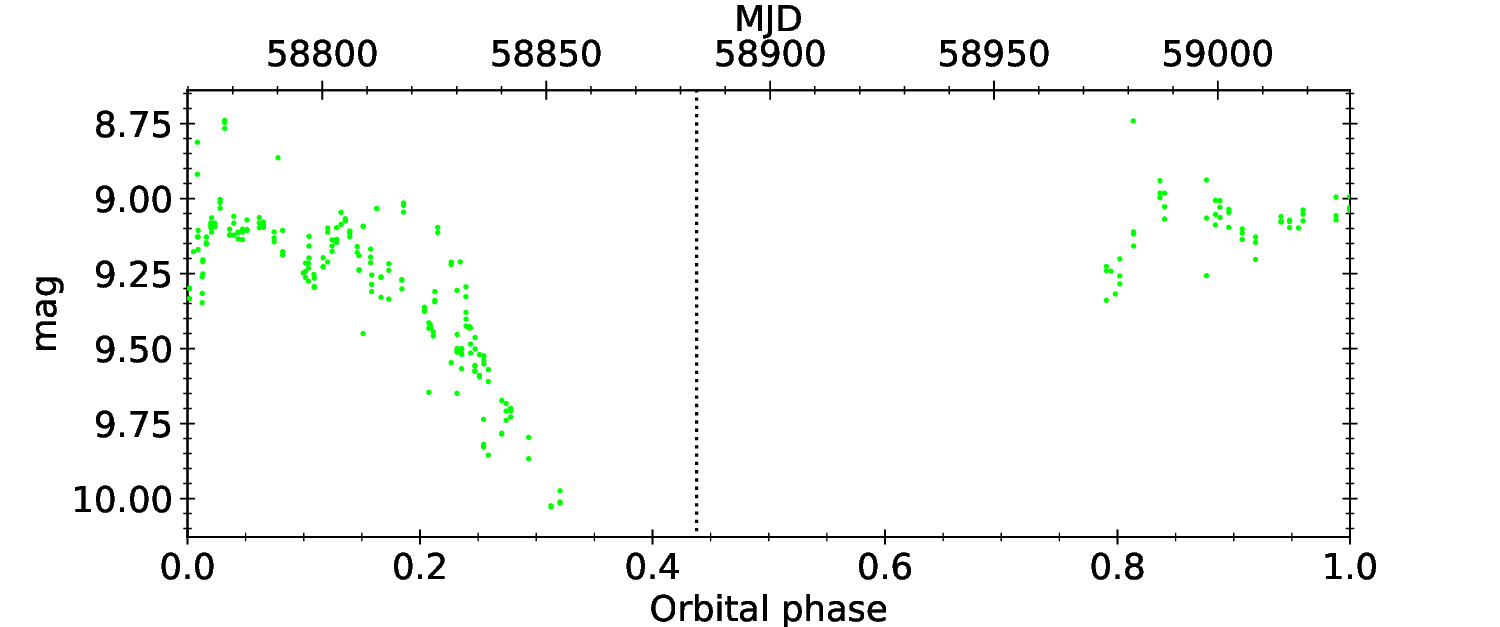}
      \caption{Continued.}
   \end{figure}

   \begin{figure}
   \ContinuedFloat
   \centering
   \includegraphics[width=\columnwidth]{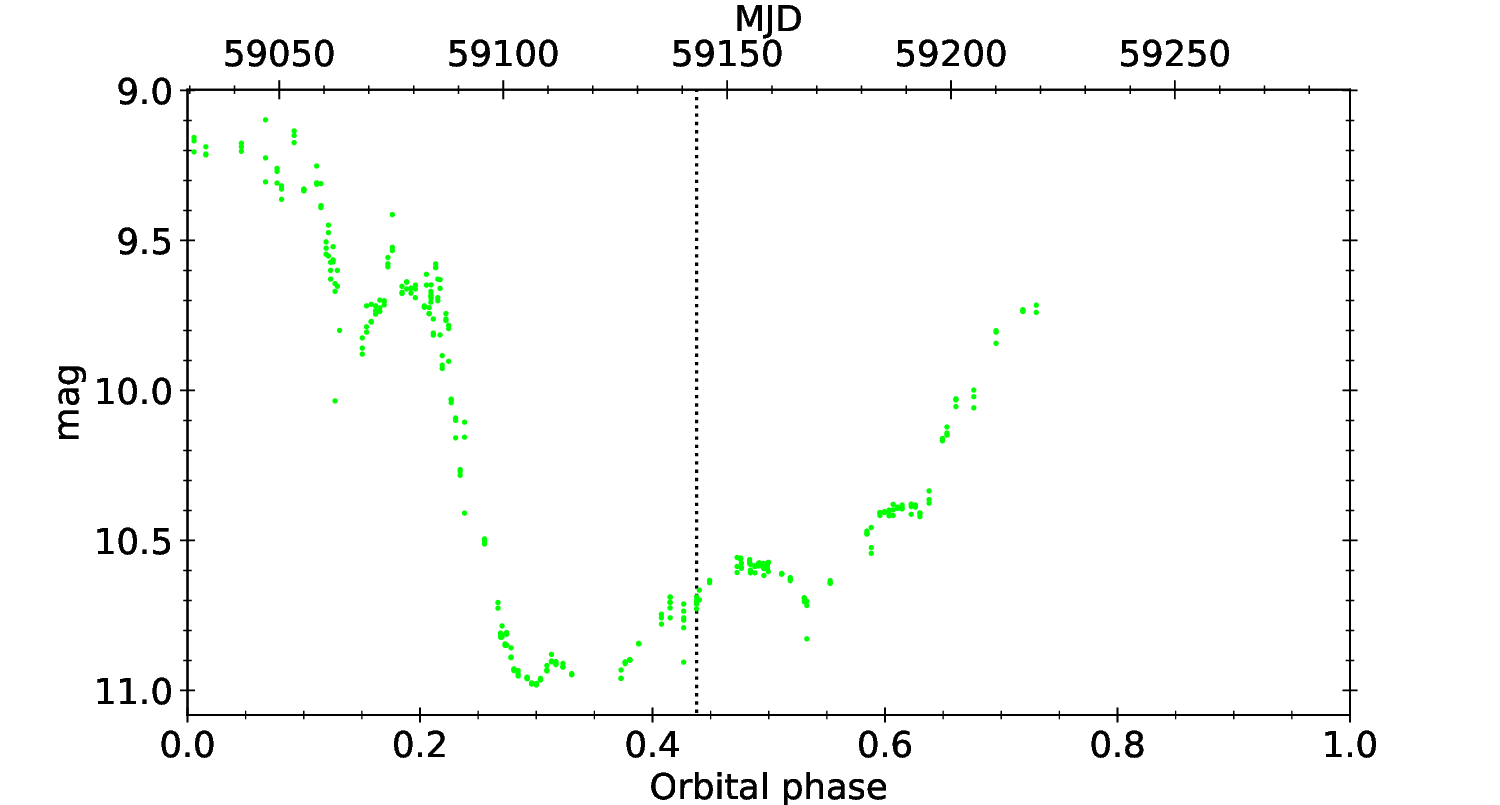}
   \includegraphics[width=\columnwidth]{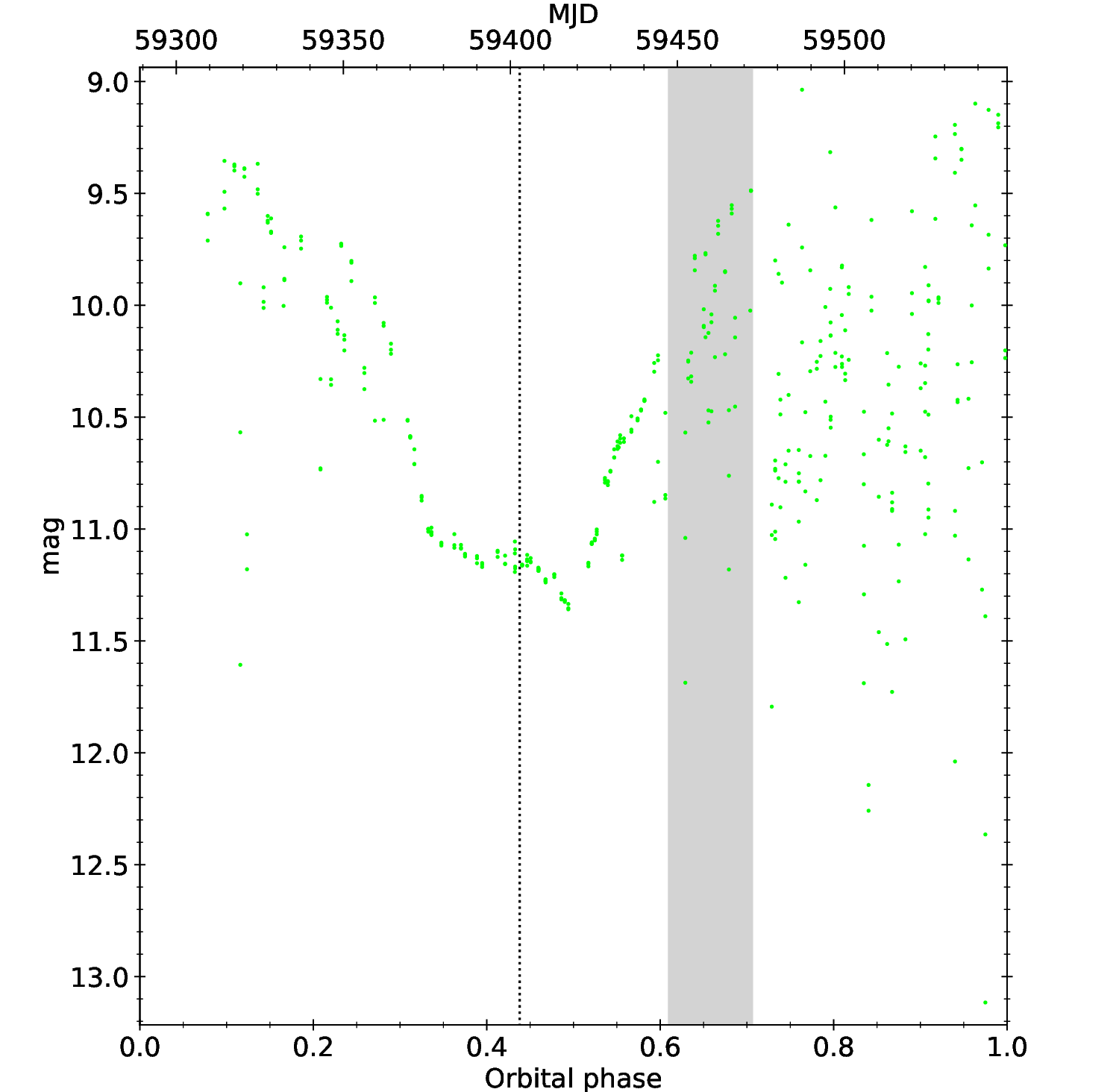}
      \caption{Continued.}
   \end{figure}

\end{appendix}

\end{document}